\documentclass{jpsj2}
\catcode`\@=11
\def\simle{\mathrel{\mathpalette\@versim<}}   
\def\simge{\mathrel{\mathpalette\@versim>}}   
\def\@versim#1#2{\lower2.5pt\vbox{\baselineskip0pt \lineskip-.5pt
   \ialign{$\m@th#1\hfil##\hfil$\crcr#2\crcr\sim\crcr}}}
\catcode`\@=12

\title{Spin Singlet State in Heptamers Emerging in Spinel Oxide AlV$_2$O$_4$}

\author{
Keisuke \textsc{Matsuda}\thanks{E-mail address: corpse@phys.aoyama.ac.jp},
Nobuo \textsc{Furukawa} and 
Yukitoshi \textsc{Motome}$^{1}$
}
\inst{
Department of Physics and Mathematics, Aoyama Gakuin University, 
Fuchinobe 5-10-1, Sagamihara, Kanagawa 229-8558 
\\
$^{1}$
Department of Applied Physics, University of Tokyo,
Hongo 7-3-1, Bunkyo-ku, Tokyo 113-8656
}

\abst{
We present our theoretical results on 
the electronic state
in vanadium spinel oxide AlV$_2$O$_4$. 
The material is a mixed-valent system
with the average valence V$^{2.5+}$, and
V cations constitute a frustrated pyrochlore structure.
It shows a structural transition at $\sim 700$~K, 
leading to the formation of seven V-sites clusters --- heptamers.
We study the electronic state of the heptamer 
by explicitly taking account of orbital degree of freedom 
as well as electron correlations. 
We show that the ground state of the heptamer for realistic parameters 
becomes spin-singlet because of
strong $\sigma$-type bonding states of $t_{2g}$ orbitals. 
The temperature dependence of the magnetic susceptibility in experiments 
is naturally understood
by this singlet formation in heptamers. 
Our results indicate that in AlV$_2$O$_4$ 
orbital physics is relevant to stabilize
a cluster-type singlet state instead of 
a previously proposed charge-ordered state with valence skipping.
}

\kword{
AlV$_2$O$_4$, spinel, pyrochlore lattice, heptamer, spin singlet, 
exact diagonalization, multiorbital Hubbard model
}

\begin{document}
\maketitle

\section{Introduction}
\label{sec:intro}
Geometrical frustration in strongly correlated systems has attracted
much interest for decades.
Many well-studied examples are found in frustrated antiferromagnets, 
where frustration results in nearly degenerate ground-state manifolds
of a large number of different spin configurations. 
\cite{Diep1994,Liebmann1986}
In some cases, the degeneracy remains down to the lowest temperature
and leads to exotic phenomena such as a liquid state and a glassy state.
In general, however, nature does not favor the degeneracy, and 
tries to find a way of lifting it and to select a unique,
nondegenerate ground state.
It is very intriguing to clarify 
the mechanism of lifting the degeneracy.

\begin{figure}[tb]
 \begin{center}
  \includegraphics[width=.9\linewidth,keepaspectratio,clip]{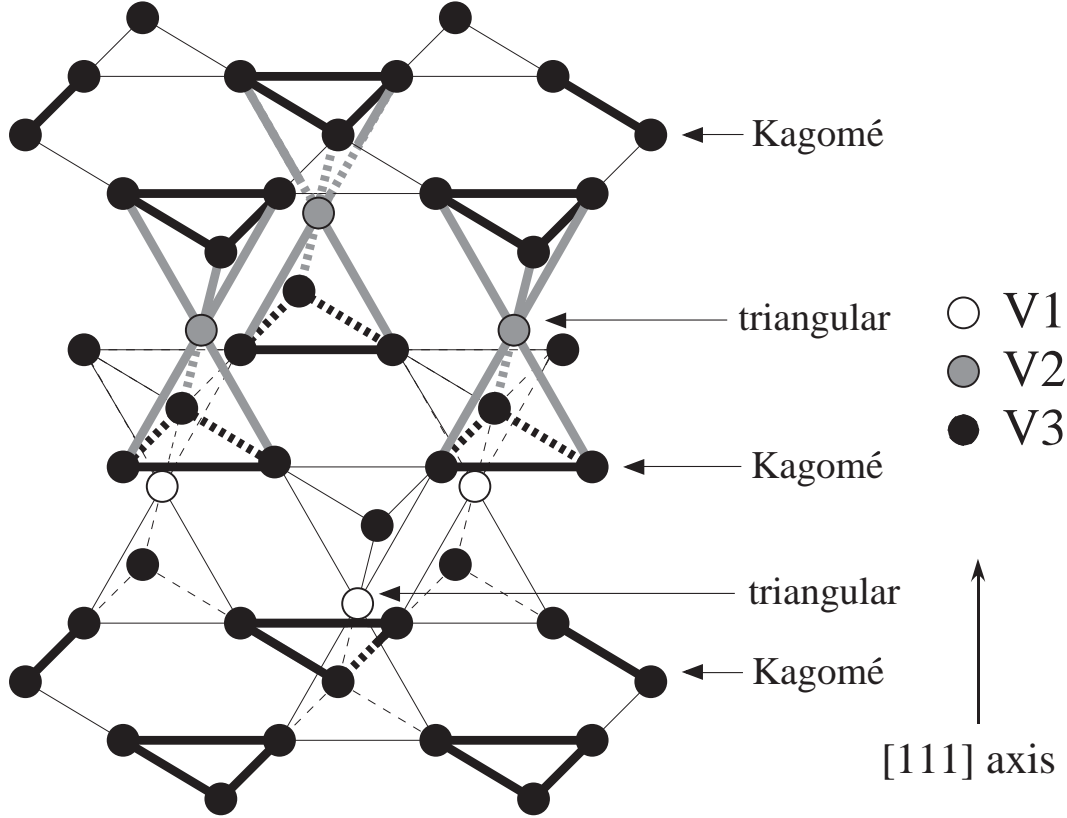}
 \end{center}
 \caption{
 Schematic picture of the lattice structure of AlV$_2$O$_4$.
 For simplicity, only V sites which constitute the pyrochlore structure are shown.
 There are three crystarographycally inequivalent V sites (V1, V2 and V3),
 and V-V bond lengths are substantially different 
 below the structural transition temperature $T_{\rm c} \sim 700$~K.
 Trimers are shown by bold lines.
 Two trimers sandwiches a V2 site (connected by gray lines),
 forming a heptamer.
 See the text for details.
 }
 \label{fig:pyrochlore}
\end{figure}

The pyrochlore lattice is a typical frustrated structure in three dimensions, and
the magnets on this lattice structure have attracted much attention
because of their severe frustration.
The pyrochlore lattice consists of a network of corner-sharing tetrahedra, 
and is regarded as a three-dimensional analogue of the Kagom\'e lattice.
In fact, the pyrochlore lattice can be viewed as
an alternative stacking of Kagom\'e and triangular planes as shown 
in Fig.~\ref{fig:pyrochlore}.
This lattice structure is found in many real compounds, for instance,
in spinels,
pyrochlore compounds
and 
cubic Laves-phase intermetallic compounds.

One of the most fundamental problems in the frustrated pyrochlore systems
was studied by Anderson.
\cite{Anderson1956}
He examined the ground state and its degeneracy 
in the Ising model on the pyrochlore lattice,
which mimics a possible charge-ordering phenomenon 
in magnetite Fe$_3$O$_4$, i.e., the so-called Verwey transition.
\cite{Verwey1939}
He pointed out that the ground state of this model 
does not exhibit any long-range order and
suffers from macroscopic degeneracy.
This shows that the frustration may give rise to nontrivial phenomena
even in charge degree of freedom.
Recently, this `charge frustration' has also been an issue  
in several other mixed-valent materials such as
CuIr$_2$S$_4$ 
\cite{Radaelli2002}
and
LiV$_2$O$_4$.
\cite{Kondo1997,Urano2000}

A spinel oxide AlV$_2$O$_4$ is a typical system 
where this charge-frustration phenomenon is addressed.
In this compound, vanadium cations form the pyrochlore lattice and 
their average valence is 2.5+.
The vanadium cations locate at the octahedral positions,
and therefore, five-fold $d$ levels of vanadium are split
by the crystal field into low-energy three-fold $t_{2g}$ levels and
high-energy two-fold $e_g$ levels.
Hence, we have 2.5 $d$ electrons in the three-fold $t_{2g}$ levels on average.
Because the $t_{2g}$ levels are partially filled,
orbital degree of freedom is also relevant in this compound 
in addition to charge and spin degrees of freedom.
Therefore, AlV$_2$O$_4$ is an interesting system to study
the effects of geometrical frustration and the roles of 
charge, spin and orbital degrees of freedom.

AlV$_2$O$_4$ shows a phase transition at $T_{\rm c} \sim$ 700 K.
At this temperature, a structural change is observed 
from the high-temperature cubic phase to the low-temperature rhombohedral phase 
accompanied by a doubling of the unit cell along the [111] direction.
\cite{Matsuno2001,Matsuno2003}
This transition was interpreted as a charge ordering
with valence skipping in which 
the charge distribution is differentiated between 
two different [111] planes, namely, 
Kagom\'e and triangular planes in Fig.~\ref{fig:pyrochlore};
it is supposed that
V cations in Kagom\'e layers take the valence 2+ and
those in triangular layers take 4+. 
However, there remain two unresolved issues in this charge-ordering scenario.
One is the discrepancy between the periodicity of the lattice structure and
the charge order.
The lattice unit cell is doubled in the [111] direction, which is
twice as long as that of charge order.
The other unresolved issue is the temperature dependence of the magnetic susceptibility.
The susceptibility shows a sudden drop at $T_{\rm c}$
followed by a Curie-like component in the low-temperature phase.
It is difficult to explain this temperature dependence on the basis of the charge-ordering scenario.

Recently, the lattice structure of this compound was reinvestigated.
\cite{Horibe2006}
It was found that below $T_{\rm c}$,
there is an additional superlattice structure
within the Kagom\'e planes, i.e., a `trimer' formation
as shown in Fig.~\ref{fig:pyrochlore}.
The trimers are paired between neighboring Kagom\'e planes,
which leads to the doubling of the unit cell along the [111] direction.
As a result, 
there are three crystallographically inequivalent V sites
as shown in Fig.~\ref{fig:pyrochlore}.
Bond lengths between V sites are estimated as 
3.04, 2.81, 2.61 and 3.14 $\AA$
for V1-V3, V2-V3, short V3-V3 (within trimers)
and long V3-V3 (between trimers) bonds, respectively.
\cite{Horibe2006}

These substantial differences among the bond lengths imply
a decomposition of the whole pyrochlore lattice into two components;
one is `heptamer', i.e., a cluster with seven V sites 
(one V2 site and six V3 sites) connected by bold and gray lines 
in Fig.~\ref{fig:pyrochlore},
and the other is
a remaining V1 site.
This decomposition and the heptamer formation may provide
an alternative scenario to the previous charge-ordering one, 
which we call the heptamer scenario hereafter.
The formation of the heptamer itself naturally explains 
the doubling of the unit cell along the [111] direction.
The peculiar temperature dependence of the susceptibility may also be compromised
if one assumes 
spin-singlet states in the heptamers and local moments at each V1 sites;
the former leads to a sudden drop at $T_{\rm c}$ and 
the latter gives rise to the Curie-like component below $T_{\rm c}$.
In fact, it was pointed out that the latter Curie-like component 
is well fitted by assuming
an $S=1$ local moment at each V1 site.
\cite{Horibe2006}

From a theoretical point of view, 
two important issues must be addressed
to understand
the physical properties of AlV$_2$O$_4$ 
on the basis of the heptamer scenario.
(i) Why are the heptamers stabilized in this frustrated pyrochlore system?
(ii) How does the spin-singlet state emerge in each heptamer?
To answer the first question, it is necessary to consider 
charge, spin and orbital degrees of freedom in the entire pyrochlore lattice, 
which is very complicated.
The second question is also nontrivial. 
There are in total 18 $d$ electrons in each heptamer
when we assign 2 $d$ electrons to each V1 site to form the $S=1$ moment
for the Curie-like contribution.
It is not trivial whether we obtain a spin-singlet state 
with 18 electrons on the seven-site cluster 
when we take account of the charge, spin and orbital states
as well as electron correlations.

In this paper, we will investigate the second problem, that is, 
the mechanism of spin-singlet formation, 
by studying the electronic state of one heptamer explicitly.
We will discuss the effects of orbital-dependent transfer integrals,
electronic correlations and the crystal field,
and clarify the parameter region where the spin-singlet ground state is obtained.
This systematic study provides a simple physical picture of
the spin-singlet formation in the heptamer.
We believe that the stabilization mechanism of
the spin-singlet state 
may give a hint 
for the first question as well.

This paper is organized as follows. 
In \S \ref{sec:hmlt},
we introduce a heptamer Hamiltonian on the basis of a multiorbital Hubbard model.
Two different singlet states in the noninteracting case 
are discussed in \S \ref{sec:scenario}. 
In \S \ref{sec:effects}, we show our numerical results
for the electronic state of the heptamer Hamiltonian. 
Section~\ref{sec:summary} is devoted for the summary and concluding remarks.

\section{Heptamer Hamiltonian}
\label{sec:hmlt}
\begin{figure}[tb]
 \begin{center}
  \includegraphics[width=.9\linewidth,keepaspectratio,clip]{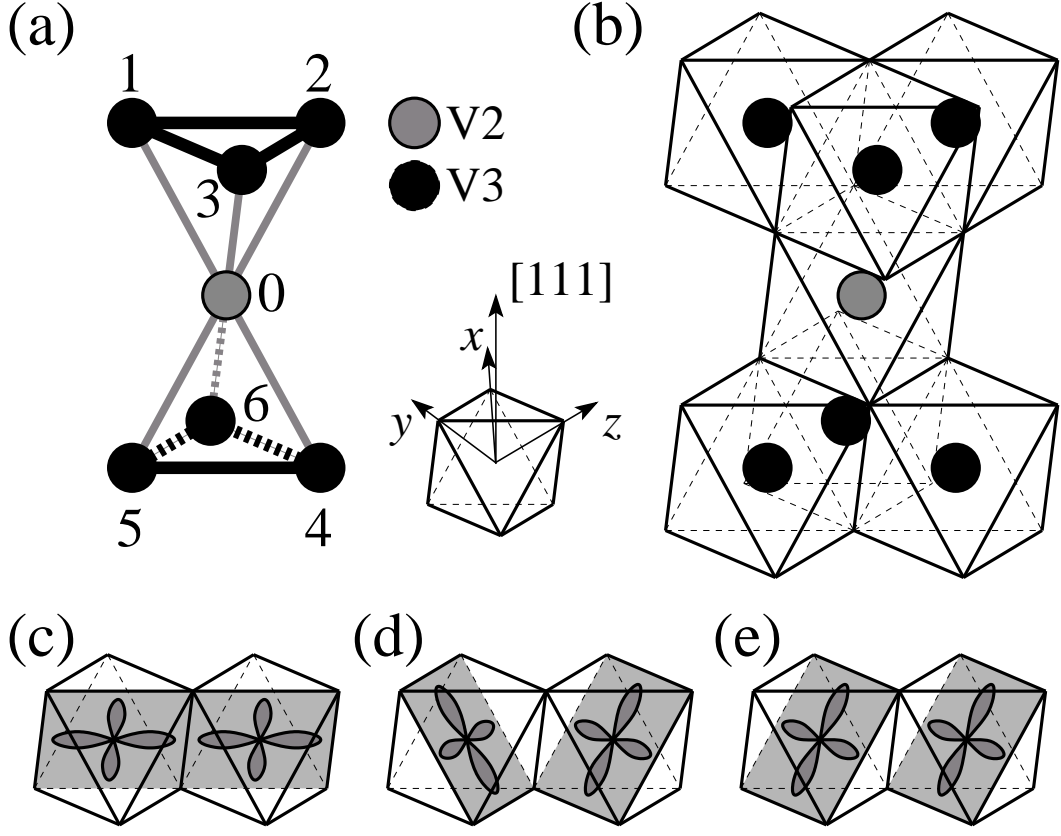}
 \end{center}
 \caption{
 (a) One heptamer. The site numbering used in our model is shown.
 (b) Edge-sharing network of VO$_6$ octahedra for the heptamer. 
 Oxygen atoms are at the corners of octahedra.
 Transfer integrals between $t_{2g}$ orbitals of 
 (c) $\sigma$-, (d) $\pi_1$- and (e) $\pi_2$-type transfer integrals, respectively.
 }
 \label{fig:hopping}
\end{figure}

To consider the electronic state of one heptamer,
we start from a multiorbital Hubbard model
with three-fold $t_{2g}$ orbital degeneracy.
The Hamiltonian consists of three terms as
\begin{align}
 \mathcal{H}_{\rm full} = 
  \mathcal{H}_{\rm kin} + \mathcal{H}_ {\rm int}
 + \mathcal{H}_{\rm trig},
 \label{eq:H_full}
\end{align}
where $\mathcal{H}_{\rm kin}$ is
the kinetic term determined by electron hoppings between V sites,
$\mathcal{H}_{\rm int}$ represents 
the on-site Coulomb interactions, and
$\mathcal{H}_{\rm trig}$ is for
the crystal-field splitting of $t_{2g}$ levels
induced by the trigonal distortion of VO$_6$ octahedra.
The Hamiltonian is defined on the seven-site cluster 
in Fig.~\ref{fig:hopping}(a), and
contains 18 electrons in total as mentioned in \S \ref{sec:intro}.

The first term $\mathcal{H}_{\rm kin}$ is given in the form
\begin{equation}
\mathcal{H}_{\rm kin}  =
 - \sum_{\langle ij \rangle} \sum_{\gamma, \gamma ' } \sum_\tau
 t_{ij}^{\gamma  \gamma '}
 \left(
 c_{i \gamma \tau}^\dagger c_{j \gamma ' \tau}
 + {\rm H.c.}
 \right),
 \label{eq:H_full_hop}
\end{equation}
where 
$i, j$ and $\tau (= \uparrow ,\downarrow) $ are site and spin indices, respectively, and
$\gamma, \gamma' = 1$ ($d_{xy}$), 2 ($d_{yz}$) and 3 ($d_{zx}$)
are orbital indices.
The site indices take the values from 0 to 6 
as shown in Fig.~\ref{fig:hopping}(a), and
the summation $\langle ij \rangle$ is taken over the nearest-neighbor sites.
(At this stage, we neglect the difference between the bond lengths
of V2-V3 bonds and V3-V3 bonds in the heptamer.)
The heptamer originally consists of the edge-sharing network of
VO$_6$ octahedra as shown in Fig.~\ref{fig:hopping}(b). 
For the edge-sharing configuration, 
the $d$-$d$ transfer integrals $t_{ij}^{\gamma \gamma'}$ 
have three different elements
as shown in Figs.~\ref{fig:hopping}(c)-(e).
Here,
we call them the $\sigma$-, $\pi_1$- and $\pi_2$-type transfer integrals.

The Coulomb interaction term $\mathcal{H}_{\rm int}$ consists of 
the four components as
\begin{align}
 \mathcal{H}_{\rm int}
 & =
 \mathcal{H}_U
 +\mathcal{H}_{U'}
 +\mathcal{H}_J
 +\mathcal{H}_{J'},
 \label{eq:H_full_int}
 \intertext{where}
 \mathcal{H}_U
 & =
 U
 \sum_i 
 \sum_\gamma 
 n_{i\gamma\uparrow} n_{i\gamma\downarrow},
 \\
 \mathcal{H}_{U'}
 & =
 U'
 \sum_i 
 \sum_{\gamma>\gamma '}
 n_{i\gamma} n_{i\gamma '},
 \\
 \mathcal{H}_J
 & =
 J
 \sum_i 
 \sum_{\gamma>\gamma '} \sum_{\tau,\tau '}
 c^\dagger_{i\gamma\tau} c_{i\gamma '\tau}
 c^\dagger_{i\gamma'\tau '} c_{i\gamma\tau '},
 \\
 \mathcal{H}_{J'}
 & =
 J'
 \sum_i 
 \sum_{\gamma \neq \gamma '}
 c^\dagger_{i\gamma\uparrow} c_{i\gamma '\uparrow}
 c^\dagger_{i\gamma\downarrow} c_{i\gamma '\downarrow}.
\end{align}
Here, $\mathcal{H}_U$, $\mathcal{H}_{U'}$, 
$\mathcal{H}_J$ and $\mathcal{H}_{J'}$ denote
the intra- and inter-orbital Coulomb interactions, 
the exchange interaction and the pair-hopping term, 
respectively.
The density operators are defined by 
$n_{i \gamma \tau} = c_{i \gamma \tau}^\dagger c_{i \gamma \tau}$
and $n_{i \gamma} = \sum_\tau n_{i \gamma \tau}$.

The third term in eq.~(\ref{eq:H_full}), $\mathcal{H}_{\rm trig}$,
represents the effect of the trigonal distortion of VO$_6$ octahedra.
The lattice structure determined by experiments indicates that
the VO$_6$ octahedron including the V2 site is 
substantially compressed in the [111] direction.
\cite{Horibe2006}
The trigonal distortion leads to the crystal-field splitting
of three-fold $t_{2g}$ levels at the V2 site
into the low-energy $a_{1g}$ singlet and the high-energy $e_g$ doublet.
This crystal-field effect is described by
\begin{equation}
 \mathcal{H}_{\rm trig} =
 -\frac{D}{2}
 \sum_{\gamma>\gamma '} \sum_\tau
 \left(
 c^\dagger_{0 \gamma\tau} c_{0 \gamma ' \tau }+{\rm H.c.}
 \right).
 \label{eq:H_full_trig}
\end{equation}
The compression of VO$_6$ corresponds to a positive $D$, 
and the energy splitting between 
the $a_{1g}$ singlet and the $e_g$ doublet is $3D/2$ in this definition.
We neglect small distortions of octahedra including V3 sites for simplicity.

The multiorbital Hubbard model given by eq.~(\ref{eq:H_full}) is too complicated
to fully handle its electronic state,
but we can make a simplification 
on the basis of the lattice structure determined by the experiment.
The key observation is that the length of the V3-V3 bonds 
is considerably shorter than that of V2-V3 bonds.
In fact, the V3-V3 bonds in the heptamer are the shortest ones
in the original pyrochlore system, 
constituting the trimers in the Kagom\'e layers.
Hence, 
we expect that the $\sigma$-type transfer integrals for these V3-V3 bonds
are the most dominant among various contributions in $t_{ij}^{\gamma \gamma'}$.
From this observation, we assume that
the $\sigma$-type bonding state for each V3-V3 bond has a sufficiently low energy 
and is fully occupied by two $d$ electrons per bonding state,
forming a singlet.
For example, the site $i=1$ is on two V3-V3 bonds
$(i,j) = (1,2)$ and $(1,3)$ 
which lie in the $yz$ and $zx$ planes, respectively, and therefore,
the $d_{yz}$ and $d_{zx}$ orbitals
constitute the bonding levels. 
This is indeed the case in the noninteracting limit 
$\mathcal{H}_{\rm int} = 0$, and is supposed to be valid
in the weak-correlation regime.
The interorbital interactions at V3 sites will be taken 
into account within the Hartree approximation
[see eq.~(\ref{eq:H_int^V3})].

The assumption of the dominant $\sigma$-type V3-V3 bonds 
allows us to simplify the model defined by eq. (\ref{eq:H_full})
in the following points.
(i) Since there are six V3-V3 bonds in the heptamer,
$6 \times 2 = 12$ electrons
occupy the $\sigma$-type bonding states.
These 12 electrons become inactive on the $\sigma$-type V3-V3 bonds.
Because we have in a total of 18 electrons in the model given by eq.~(\ref{eq:H_full}),
we will consider the remaining 6 electrons in the following calculations.
Hereafter,
we call the orbitals constituting the $\sigma$-type V3-V3 bonds inactive orbitals and 
the remaining orbitals active orbitals.
(ii) All $\pi_2$-type transfer integrals
[Fig.~\ref{fig:hopping}(e)] can be neglected,
because 
orbitals which are connected to active orbitals through $\pi_2$-type bonds are always inactive.
(iii) $\pi_1$-type transfer integrals for V2-V3 bonds can also be neglected
on the basis of a reason similar to that in item (ii).
As a consequence of these simplifications, among the transfer integrals,
only two contributions are relevant;
one is the $\sigma$-type transfer integral for V2-V3 bonds,
which is denoted by $t_\text{V2-V3}^\sigma$, 
and the other is the $\pi_1$-type one for V3-V3 bonds, 
$t_\text{V3-V3}^\pi$. 

Consequently,
on the basis of the above assumption,
the Hamiltonian given by eq.~(\ref{eq:H_full}) is simplified into the form
\begin{equation}
\mathcal{H} = \tilde{\mathcal{H}}_{\rm kin}
+ \tilde{\mathcal{H}}_{\rm int} + \mathcal{H}_{\rm trig}.
\label{eq:H}
\end{equation}
The kinetic term for the remaining 6 electrons, 
$\tilde{\mathcal{H}}_{\rm kin}$, is given by
\begin{equation}
\tilde{\mathcal{H}}_{\rm kin} = \mathcal{H}_{\rm kin}^\text{V2-V3} + 
\mathcal{H}_{\rm kin}^\text{V3-V3},
\end{equation}
where $\mathcal{H}_{\rm kin}^\text{V2-V3}$ denotes
the $\sigma$-type hoppings between the V2 site ($i=0$) and
V3 sites ($i=1-6$), and
$\mathcal{H}_{\rm kin}^\text{V3-V3}$ denotes
the $\pi_1$-type hoppings between the nearest-neighbor V3 sites:
\begin{align}
\mathcal{H}_{\rm kin}^\text{V2-V3} & = - t_\text{V2-V3}^\sigma
\sum_{i=1}^6 
\sum_\tau \left( c_{0 \alpha_i \tau}^\dagger
c_{i \alpha_i \tau} + {\rm H.c} \right),
\label{eq:H_hop_V2-V3}
\\
\mathcal{H}_{\rm kin}^\text{V3-V3} &= - t_\text{V3-V3}^\pi
\sum_{\langle ij \rangle \in \text{V3}} \sum_\tau
\left( c_{i \alpha_i \tau}^\dagger c_{j \alpha_j \tau} 
+ {\rm H.c.} \right).
\label{eq:H_hop_V3-V3}
\end{align}
Here,
$\alpha_i$ denotes 
the active orbital at the $i$-th V3 sites
[for instance, $\alpha_1 = 1$ ($d_{xy}$)].
Note that in eq.~(\ref{eq:H_hop_V2-V3}), 
the orbital $\alpha_i$ constitutes 
the $\sigma$-type bond between 0-th and $i$-th sites, 
while the orbitals $\alpha_i$ and $\alpha_j$ at neighboring V3 sites
constitute the $\pi_1$-type bond in eq.~(\ref{eq:H_hop_V3-V3})
where the summation 
is taken over the nearest-neighbor V3 sites.
The Coulomb interaction term in eq.~(\ref{eq:H}) is written as
\begin{equation}
 \tilde{\mathcal{H}}_{\rm int} =
  \mathcal{H}_{\rm int}^{\rm V2} + \mathcal{H}_{\rm int}^{\rm V3}.
\end{equation}
The former term is for the V2 site, given by the terms 
in eq.~(\ref{eq:H_full_int}) for $i=0$ only.
The latter is for the V3 sites,
which is reduced into
\begin{equation}
 \mathcal{H}_{\rm int}^{\rm V3} = 
  U 
  \sum_{i=1}^6 
  n_{i \alpha_i \uparrow} n_{i \alpha_i \downarrow}
  + (2U' - J) 
  \sum_{i=1}^6 
  n_{i \alpha_i},
  \label{eq:H_int^V3}
\end{equation}
where $\alpha_i$ denotes the active orbital as in eqs.~(\ref{eq:H_hop_V2-V3})
and (\ref{eq:H_hop_V3-V3}).
Here, the second term is the Hartree potential
from electrons in the $\sigma$-type bonding states assumed
for the V3-V3 bonds.

\section{Noninteracting Case: Two Different Scenarios for Singlet Formation}
\label{sec:scenario}

Before going into the numerical study of
the electronic state of the heptamer model given by eq.~(\ref{eq:H}),
here we consider the noninteracting case,
i.e., $\tilde{\mathcal{H}}_{\rm int} = 0$ ($U = U' = J = J' = 0$).
In this case, we have 
three parameters $t_\text{V2-V3}^\sigma$,
$t_\text{V3-V3}^\pi$ and $D$. 

\begin{figure}[tb]
 \begin{center}
  \includegraphics[width=.8\linewidth,keepaspectratio,clip]{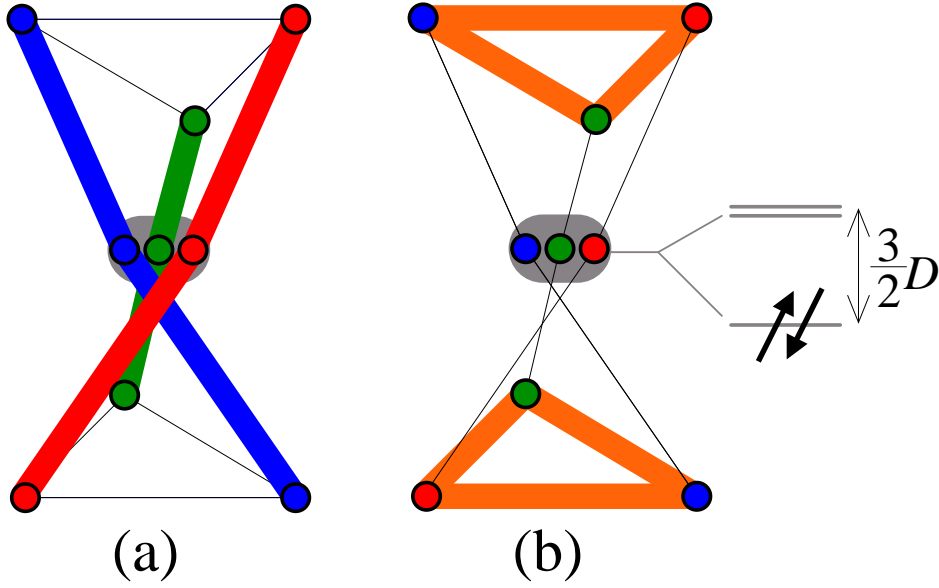}
 \end{center}
 \caption{
 Schematic pictures of two different singlet states
 in the noninteracting case, 
 (a) $\sigma$-singlet state [eq.~(\ref{eq:sigma wf})] and 
 (b) $\pi$-singlet state [eq.~(\ref{eq:pi wf})].
 Blue, red and green circles represent $d_{xy}$, $d_{yz}$
 and $d_{zx}$ orbitals at each V site, respectively.
 Singlet bonds are shown by bold lines.
 In (b), at the V2 site, two electrons occupy the $a_{1g}$ singlet state
 due to the trigonal splitting.
 }
 \label{fig:scenario}
\end{figure}

There are two important cases in this parameter space
to consider the singlet formation in the heptamer.
One is $t_\text{V2-V3}^\sigma \neq 0$ and $t_\text{V3-V3}^\pi = D = 0$.
The $\sigma$-type transfer integrals in an orbital sector (say $d_{xy}$)
have finite matrix elements on one straight V3-V2-V3 bond
lying in the corresponding plane (in the $xy$ plane).
Hence, in this case, three-site clusters are formed 
on straight V3-V2-V3 bonds in each orbital sector,
for instance, a cluster with the sites $i=1, 0, 4$ 
in the $d_{xy}$ orbital sector.
In each cluster, the ground state is given by the bonding state
of three sites, being singlet as shown in Fig.~\ref{fig:scenario}(a).
The ground-state wave function for the heptamer is given by
\begin{equation}
 |\sigma \rangle =
  \prod_\tau \prod_{i=1}^3
  \frac12 \left( c_{i \alpha_i \tau}^\dagger +
	   \sqrt{2} c_{0 \alpha_i \tau}^\dagger +
	   c_{(i+3) \alpha_i \tau}^\dagger \right) |0 \rangle,
  \label{eq:sigma wf}
\end{equation}
where $|0 \rangle$ is the vacuum state of the model in eq.~(\ref{eq:H}).
We call this the $\sigma$-singlet state.
Note that the state in eq.~(\ref{eq:sigma wf}) is obtained 
for both positive and negative $t_\text{V2-V3}^\sigma$.

The other important case is $t_\text{V2-V3}^\sigma = 0$
and $t_\text{V3-V3}^\pi, D > 0$.
In this case, the heptamer is separated into 
the isolated V2 site and two trimers with V3 sites.
In each trimer, the ground state is given by the bonding state
due to the $\pi_1$-type transfer integrals, $t_\text{V3-V3}^\pi$.
At the V2 site, for a positive $D$, the ground state is
given by a doubly occupied $a_{1g}$ singlet
as shown in Fig.~\ref{fig:scenario}(b).
Hence, the wave function for the heptamer is expressed as
\begin{equation}
 |\pi \rangle =
  \prod_\tau 
  \sum_{\gamma=1}^3 \frac{c_{0 \gamma \tau}^\dagger}{\sqrt{3}}
  \sum_{i=1}^3 \frac{c_{i \alpha_i \tau}^\dagger}{\sqrt{3}}
  \sum_{i=4}^6 \frac{c_{i \alpha_i \tau}^\dagger}{\sqrt{3}}
  |0 \rangle.
  \label{eq:pi wf}
\end{equation}
This state is schematically shown in Fig.~\ref{fig:scenario}(b). 
We call this the $\pi$-singlet state.
Note that  
if $t_\text{V3-V3}^\pi$ or $D$ is negative, 
the ground state no longer becomes singlet and has some degeneracy.

When we switch on the interactions, 
it is nontrivial that the ground state is singlet.
Nevertheless, as we will see in the numerical results in the next section,
there appear two different singlet regions;
one includes the $\sigma$-singlet limit and
the other includes the $\pi$-singlet one.
In other words, in each region, the wave function is 
adiabatically connected with either $|\sigma \rangle$ or $|\pi \rangle$.
Besides these two singlet regions,
there appears also a doublet nonmagnetic ground state.
Thus, our questions are twofold.
(i) For realistic values of parameters,
does the ground state of the heptamer Hamiltonian become singlet?
(ii) If that is the case, which category does the singlet state belong to,
the $| \sigma \rangle $- or $| \pi \rangle $-like state?

\section{Effects of Electron Correlation: Exact Diagonalization Study}
\label{sec:effects}
\subsection{Method and parameters}
\label{sec:pm}
We investigate the electronic state of the model in eq.~(\ref{eq:H})
including the interaction terms
by exact diagonalization (ED).
We use the Householder method for the ED calculations.

For the $d$-$d$ transfer integrals, 
the Slater-Koster scheme
\cite{Slater1954}
gives the estimates 
$t_\text{V2-V3}^\sigma \simeq 0.5$~eV and
$t_\text{V3-V3}^\pi \simeq -0.26$~eV
on the basis of using the bond lengths in experiments.
Hereafter, we set $t_\text{V2-V3}^\sigma = 1$ as an energy unit, 
and change $t_\text{V3-V3}^\pi$ from $-2$ to $2$ 
and $D$ from $0$ to $2$ 
to study overall features in the parameter space,
which is supposed to include realistic values for the present $t_{2g}$ system.

Concerning the interaction parameters, 
no experimental estimate for AlV$_2$O$_4$ is available yet. 
An estimate was given for a related spinel vanadate LiV$_2$O$_4$;
\cite{Fujimori1988}
the optical measurement suggests $U \simge 2$eV.
In the following, we present the results for $U = 0-8$
to show the tendency with changing $U$.
We set $J=0.1U$, and retain the relations $U=U'+2J$ and $J=J'$
\cite{Kanamori1963}.

\subsection{Results}
\label{sec:ed}
\begin{figure}[tb]
 \begin{center}
 \includegraphics[width=0.7\linewidth,keepaspectratio,clip]{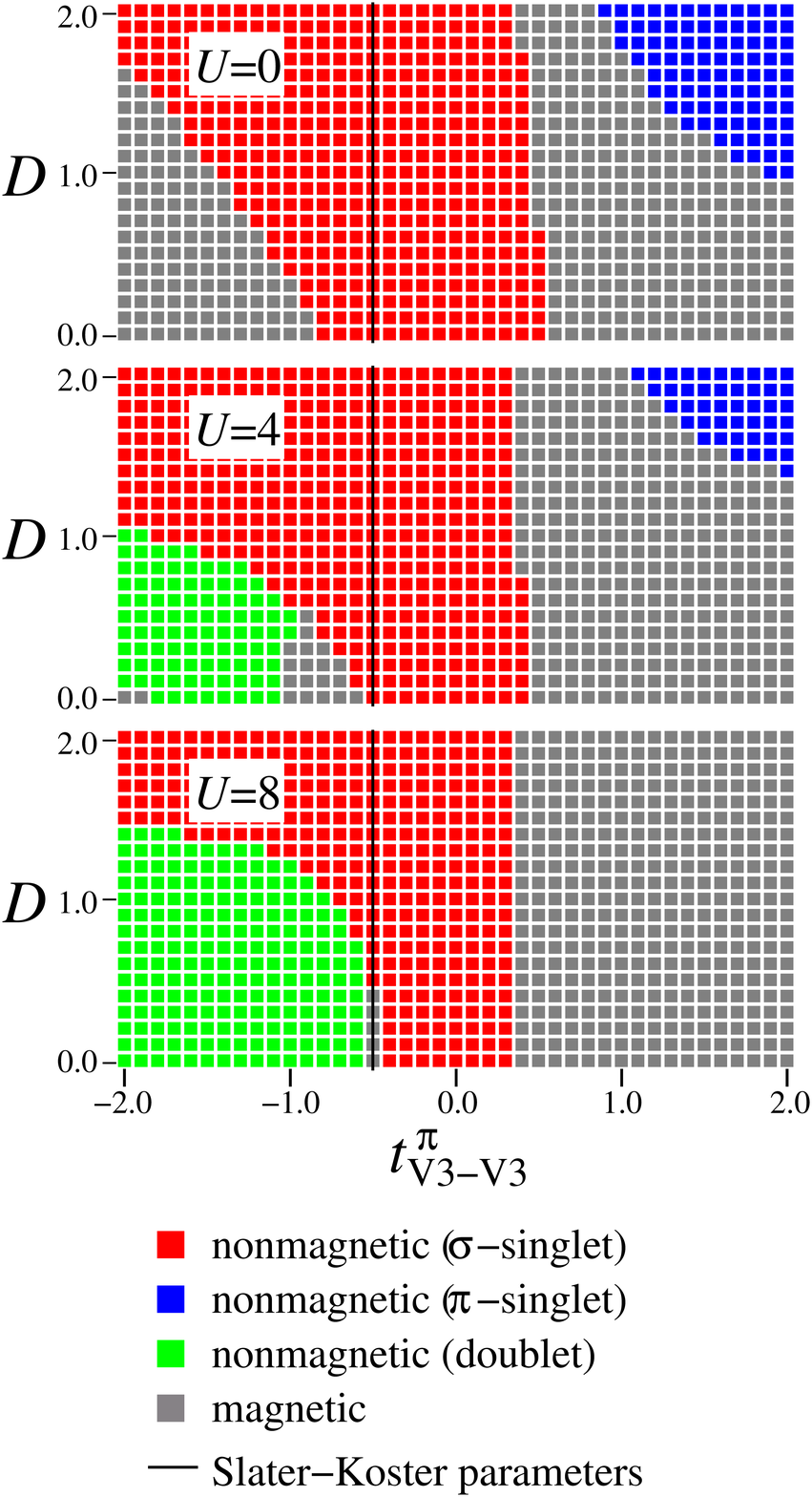}
 \caption{Magnetism of the ground state of the heptamer model for several $U$.
 The red, blue and green regions denote the $\sigma$, $\pi$-singlet and doublet regions, respectively.
 See the text for details.}
 \label{fig:phase}
 \end{center}
\end{figure}

First,
we study the magnetism and the degeneracy of the ground state of 
the heptamer Hamiltonian given by eq.~(\ref{eq:H}).
Figure~\ref{fig:phase} shows 
the results
in the $t_\text{V3-V3}^\pi$--$D$ plane for different values of $U$.
Nonmagnetic ground states are obtained in three regions in the present parameter space;
two singlet regions without any degeneracy (red and blue) and
a doublet region (green).
Two singlet regions are well separated, and
each region includes one of the limits in the noninteracting case
discussed in \S 3;
the red region 
extends from the $\sigma$-singlet limit ($t_\text{V3-V3}^\pi = D = U = 0$),
the blue region 
continues to the $\pi$-singlet limit ($t_\text{V2-V3}^\sigma = U =0$ and $t_\text{V3-V3}^\pi, D>0$).
The doublet region
emerges only
for finite values of $U$.
Both singlet regions shrink as $U$ increases, 
but remain well for intermediate or rather large values of $U$,
while
the doublet region spreads as $U$ increases.

\begin{figure}[tb]
 \begin{center}
 \includegraphics[width=0.8\linewidth,keepaspectratio,clip]{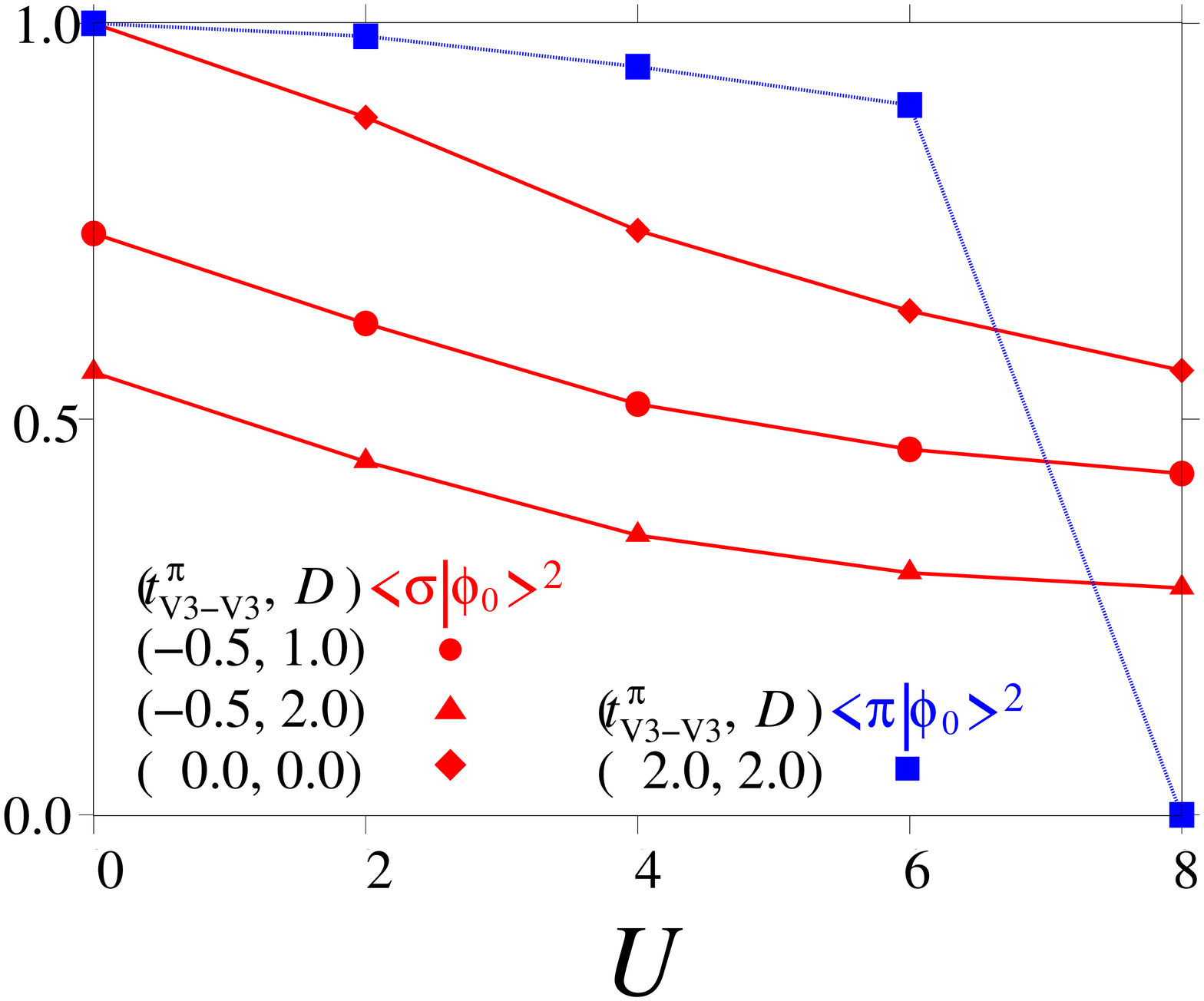}
 \caption{
 Square of overlap $ \langle \sigma | \phi_0 \rangle ^2 $ and $ \langle \pi | \phi_0 \rangle ^2 $
 for several sets of $(t_\text{V3-V3}^\pi,D)$.
 The solid lines are guides for the eyes.
 }
 \label{fig:U_overlap}
 \end{center}
\end{figure}%

We focus on two singlet regions, one of which is believe to be relevant
to understanding the experiments in AlV$_2$O$_4$ as discussed below.
We will discuss the doublet region later.
Although these two singlet regions are separated in the parameter space,
the ground states for two regions show the same symmetry.
The symmetry of the ground state of the model given by eq.~(\ref{eq:H}) belongs to 
one of the irreducible representations $A_{1}$, $A_{2}$ and $E$
because the Hamiltonian belongs to the $D_3$ point group.
\cite{Altmann1994}
We calculate
$\langle \phi_0 | C_3 | \phi_0 \rangle$ and 
$\langle \phi_0 | C'_2 | \phi_0 \rangle$,
where $C_3$ and $C'_2$ are operators for the symmetric rotation 
of a threefold symmetric axis by $3\pi / 2$ and that 
of a twofold symmetric axis by $\pi$, respectively.
Here, $| \phi_0 \rangle$ is the ground-state wave vector obtained by ED calculations.
We found that 
$\langle \phi_0 | C_3 | \phi_0 \rangle = 
\langle \phi_0 | C'_2 | \phi_0 \rangle = 1$ 
for all the ED solutions without degeneracy.
Namely, all the ground states in both separated singlet regions 
belong to the $A_1$ symmetry group.

To characterize two singlet regions,
we calculate overlaps of the ground-state wave function
with $|\sigma \rangle$ and $|\pi \rangle$ 
in eqs.~(\ref{eq:sigma wf}) and (\ref{eq:pi wf}), respectively.
We plot the square of overlaps in Fig.~\ref{fig:U_overlap}
for typical sets of parameters. 
In the red region in Fig.~\ref{fig:phase},
the square of the overlap with $ | \sigma \rangle $,
$ \langle \sigma | \phi_0 \rangle ^2 $, is substantial 
whereas $ \langle \pi | \phi_0 \rangle ^2 $ is negligibly small ($< 0.01$).
Since $ \langle \sigma | \phi_0 \rangle ^2 \gg \langle \pi | \phi_0 \rangle ^2 $,
we call this red region the $\sigma$-singlet region.
On the other hand,
in the blue region in Fig.~\ref{fig:phase},
$ \langle \pi | \phi_0 \rangle ^2 $ has a substantial value
compared with $ \langle \sigma | \phi_0 \rangle ^2 < 0.01$,
and therefore 
we call the blue region the $\pi$-singlet region.
Hence,
these two singlet regions are well distinguished by the values of overlaps,
and appear to be separated within the parameter space 
that we investigated.

\begin{figure}[tb]
 \begin{center}
 \includegraphics[width=0.7\linewidth,keepaspectratio,clip]{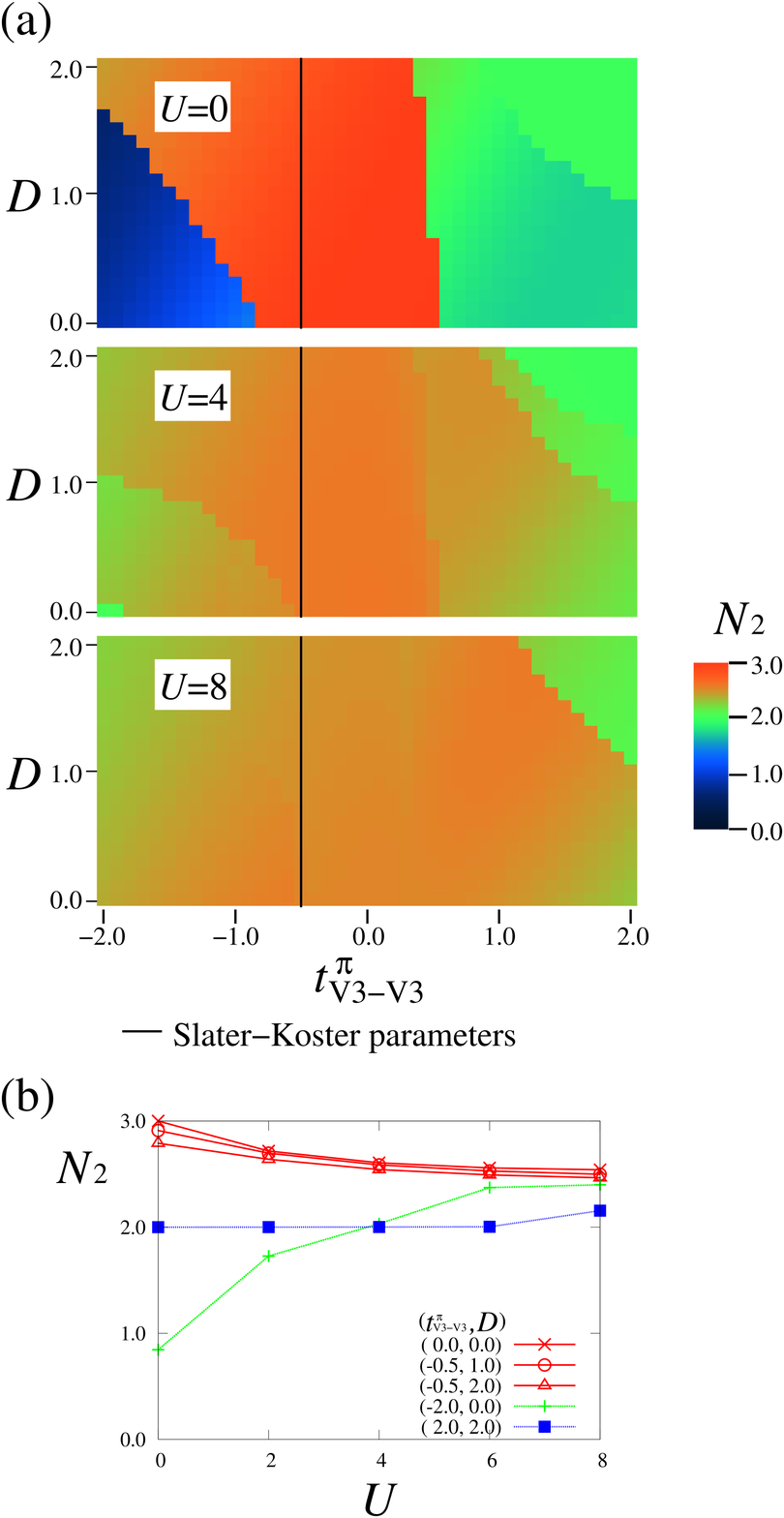}
 \caption{
 (a) Charge density at V2 site, $N_2$, for several $U$.
 In the degenerate regions,
 we plot the values averaged over all degenerate states.
 $U$ dependences are shown in (b) for several sets of $(t_\text{V3-V3}^\pi,D)$.
 The solid lines in (b) are guides for the eyes.
 }
 \label{fig:N2}
 \end{center}
\end{figure}

We also calculate the charge disproportionation between the V2 site and the V3 sites
in the ground state.
Figure~\ref{fig:N2} shows the electron density at the V2 site;
\begin{equation}
 N_2 = \langle \phi_0 |
  \sum_\gamma n_{0 \gamma} 
  | \phi_0 \rangle.
\end{equation}
Note that 
the electron density at the V3 site is given by $(18-N_2)/6$ 
for singlet states with $A_1$ symmetry.
In the $ \sigma $-singlet region (red region in Fig.~\ref{fig:phase}), as $U$ increases,
$N_2$ continuously decreases from the noninteracting value $3$. 
On the contrary, in the $\pi$-singlet region (blue region in Fig.~\ref{fig:phase}),
$N_2$ is almost unchanged and remains at the noninteracting value $2$.
This difference comes from the nature of the singlet states.
In the former, since the dominant singlet bonds are 
on the straight V3-V2-V3 clusters, 
electron charges are easily redistributed 
between V2 and V3 sites.
On the other hand, in the latter $\pi$-singlet region,
the dominant singlet bonds are confined within V3 trimers,
and the bonds between V2 and V3 sites are relatively weak.
There, the singlet energy gain within the V3 trimers as well as
the crystal field $D$ at the V2 site prevents the charge redistribution
between V2 and V3 sites.

The results in Figs.~\ref{fig:phase} and \ref{fig:U_overlap} show that
for the estimates of transfer integrals by using the Slater-Koster scheme,
the ground state of the model given by eq.~(\ref{eq:H}) is singlet
and belongs to the $\sigma$-singlet region
where the ground-state wave function is close to $| \sigma \rangle$ 
in eq.~(\ref{eq:sigma wf}).
We note that the $\pi$-singlet region is only limited for large values of
$t_\text{V2-V3}^\pi$ and $D$ compared with $t_\text{V2-V3}^\sigma$,
and this is not realistic in the present $t_{2g}$ electron system.
Therefore, we conclude that the singlet ground state is realized
in the heptamer, 
and that the basic mechanism of singlet formation is understood
on the basis of the $\sigma$-type bonding of $t_{2g}$ orbitals.

\begin{figure}[tb]
 \begin{center}
  \includegraphics[width=0.8\linewidth,keepaspectratio,clip]{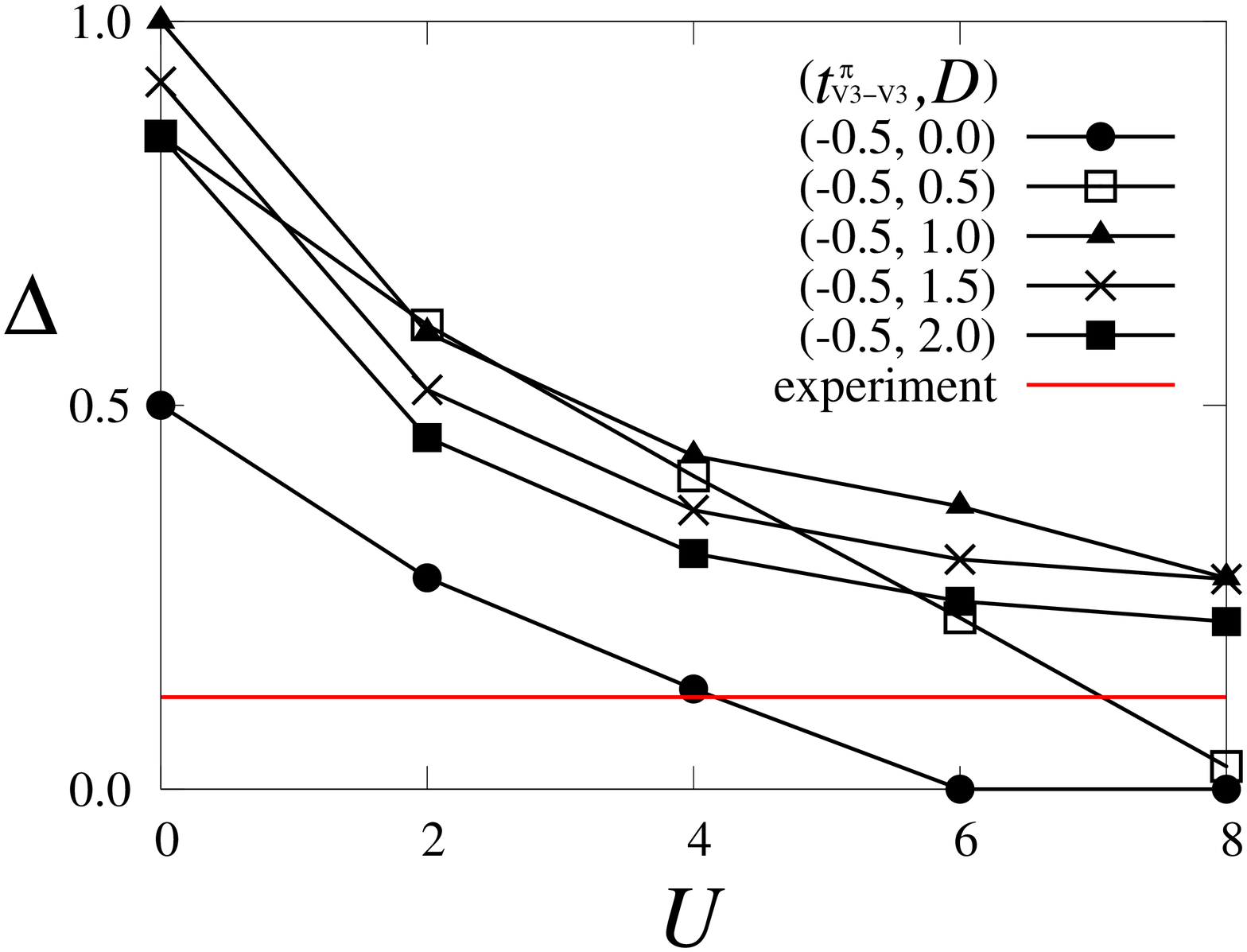}
  \caption{
  Spin gap for the heptamer model given by eq.~(\ref{eq:H})
  for several Slater-Koster parameters ($t_\text{V3-V3}^\pi ,D$). 
  }
  \label{fig:egap}
 \end{center}
\end{figure}

We calculate the spin gap $\Delta$ 
for a comparison with experimental data,
which we define here as the energy gap
from the ground state to the lowest magnetic excited state.
Note that the spin gap becomes zero when the ground state is magnetic by definition.
The result is shown in Fig.~\ref{fig:egap} for several Slater-Koster parameters.
Although a quantitative comparison with the experimental estimate
is beyond the scope of the present study 
since interactions among the heptamers are completely neglected
in the present theory,
\cite{note}
this finite spin gap gives rise to the suppression of magnetic susceptibility
below $T_c$ observed in experiments.
Therefore,
our results on the gapped, spin-singlet ground state in the heptamer model
support the heptamer picture,
in which the peculiar temperature dependence of the magnetic susceptibility
is explained as the summation of the heptamer-singlet contribution
and the $S=1$ Curie-like contribution.
\cite{Horibe2006}

Finally, let us make a remark on the doublet state (green region) 
in Fig.~\ref{fig:phase}.
The doublet region develops as $U$ increases, and includes 
the parameter regime given by the Slater-Koster scheme 
for large values of $U > 8$.
From the symmetry analysis, we find that the doublet state belongs
to $E$ symmetry.
Since this state is also nonmagnetic, 
it might be possible to explain the peculiar temperature dependence
of the magnetic susceptibility.
Nevertheless, we believe that the $\sigma$-type singlet ground state
is relevant to AlV$_2$O$_4$ for the following reasons.
One is that the degeneracy remaining in the doublet state should be
lifted by the generalized Jahn-Teller mechanism or the heptamer-heptamer interaction
when we allow a further lattice distortion of the heptamer; however,
such lowering of the symmetry is not observed in experiments.
The other reason is that our present analyses are justified
in the weak correlation regime as discussed in \S \ref{sec:hmlt}, 
and hence some effects beyond our assumptions may alter 
the results in the relatively large $U$ region.

\section{Summary and Concluding Remarks}
\label{sec:summary}
We have investigated the origin of spin-singlet formation
in seven-site clusters, heptamers, 
which emerge 
in the spinel oxide AlV$_2$O$_4$ 
as a result of the structural change 
with the trimer formation in Kagom\'e layers of the V pyrochlore lattice. 
We have derived an effective model to describe 
the charge, spin and orbital degrees of freedom in one heptamer 
from the Hubbard model with $t_{2g}$ orbital degeneracy, 
and investigated its electronic and magnetic states 
using numerical exact diagonalization. 
Our results indicate that
in the realistic parameter region for AlV$_2$O$_4$,
the ground state of the heptamer becomes spin-singlet, and that
not the trimer formation in the Kagom\'e layers
but the heptamer formation
plays an essential role in the spin-singlet formation in this system.
We conclude that
the singlet state is stabilized by dominant $d$-$d$ transfer integrals
of the $\sigma$-type 
which form bonding states within the heptamer.
The heptamer formation can give a comprehensive understanding of
the magnetic properties observed in experiments; 
a suppression due to the singlet gap formation and
the Curie-like behavior at lower temperatures 
coming from $S=1$ moments outside the heptamers. 

How can we distinguish experimentally 
the three different origins of spin-singlet formation,
i.e., two different singlet states ($\sigma$- and $\pi$-type) 
and one doublet state?
We can distinguish between singlet and doublet states
by the entropy at high temperatures 
estimated from the electronic specific heat.
Then, 
it is possible to distinguish between $\sigma$- and $\pi$-type singlet states
which have the same symmetry.
One possible way is to detect the directional dependence of
the spin excitation. 
The dominant singlet bonds lie on
the intertrimer V3-V2-V3 bonds in the $\sigma$ case and 
within the V3 trimers on Kagom\'e planes in the $\pi$ case. 
This difference may be detected 
using the Raman scattering technique. 
Another possibility is to measure the local density of $d$ electrons at the V2 site.
Our results in Fig.~\ref{fig:N2}(f) indicate that the local density 
becomes $\sim 2.5 - 3.0$ in the $\sigma$-singlet region,
while it is $\simeq 2.0$ in the $\pi$-singlet one.

Let us comment on the properties of $S=1$ moments outside the heptamers.
In our heptamer scenario,
below $T_{\rm c}$,
the magnetic moments that remain active are 
the $S=1$ spins at V1 sites only,
and the spins at V2 and V3 sites are magnetically inactive,
because of the singlet formation in heptamers.
Since a V1 site is connected to the V2 sites (see Fig.~\ref{fig:pyrochlore}),
the V1 spins are almost magnetically isolated.
At a much lower temperature than $T_{\rm c}$,
another anomaly appears at $T \sim 5$K in the magnetic susceptibility,
which is ascribed to spin-glass transition.
\cite{Matsuno2001}
This indicates that in this low-temperature region,
$S=1$ spins begin to interact with each other.
Nearest-neighbor V1 pairs correspond to the third-nearest-neighbor sites 
across hexagons in the original pyrochlore lattice, 
forming a two-dimensional triangular lattice in the [111] planes. 
Hence, the $S=1$ moments at V1 sites constitute
a weakly coupled triangular spin system at low temperatures. 
The sign of exchange interactions is not trivial
because of complicated exchange paths, but in any case, 
it is believed that Heisenberg spin systems on a weakly coupled triangular lattice 
show a magnetic order, not a spin-glass behavior. 
Therefore, our heptamer scenario suggests that 
the spin-glass behavior comes from an extrinsic effect, such as randomness.

Our present study was conducted on the basis of
the lattice structure below the structural transition temperature, 
and therefore, 
at this stage, we cannot answer why and how the lattice distortion
including the formation of heptamers is stabilized 
in the frustrated pyrochlore lattice structure. 
There are many similar examples of such clusters 
in frustrated systems, such as
hexamers in ZnCr$_2$O$_4$,
\cite{Lee2002} 
octamers in CuIr$_2$S$_4$,
\cite{Radaelli2002,Khomskii2004}
trimers in LiVO$_2$,
\cite{Pen1997} and
dodecamers in a double-exchange spin-ice model 
on a Kagom\'e lattice.
\cite{Shimomura2004}
These facts suggest the ubiquitous role 
of cluster formations 
in lifting the degeneracy inherent 
in frustrated systems,
although the driving force may depend on the details of systems. 
In the present system AlV$_2$O$_4$, 
our results imply that 
the stability of the heptamer singlet state itself 
may play some role in the mechanism of self-organizing the clusters.
However, to identify a dominant player, 
it is necessary to handle charge, spin, orbital and lattice degrees of freedom 
and to compare the instabilities to many possible ordered states 
on the frustrated pyrochlore lattice.
This is left for future study. 

\section*{Acknowledgment}
We would like to thank M. Shingu, T. Katsufuji, S. Mori and Y. Horibe 
for stimulating discussions.
This work was supported by a Grant-in-Aid for the 21st COE program 
and for Scientific Research (No. 16GS50219)
from the Ministry of Education, Culture, Sports, Science and Technology of Japan.


\begin{thebibliography}{99}
 \bibitem{Diep1994}
	 {\it Magnetic System with Competing Interaction}, 
	 ed. by H. T. Diep (World Scientific Publishing Co. Pte. Ltd., Singapore, 1994).
 \bibitem{Liebmann1986}
	 R. Liebmann, {\it Statistical Mechanics of Periodic Frustrated Ising Systems} 
	 (Springer-Verlag, Berlin, Tokyo, 1986).
 \bibitem{Anderson1956} P. W. Anderson: Phys. Rev. \textbf{102} (1956) 1008.
 \bibitem{Verwey1939} E. J. W. Verwey: Nature (London) \textbf{144} (1939) 327.
 \bibitem{Radaelli2002} P. G. Radaelli, Y. Horibe, M. J. Gutmann, H. Ishibashi, C. H. Chen, 
	 R. M. Ibberson, Y. Koyama, Y. S. Hor, V. Kiryukhin and S-W. Cheong:
	 Nature \textbf{416} (2002) 155.
 \bibitem{Kondo1997} S. Kondo, D. C. Johnston, C. A. Swenson, F. Borsa, A. V. Mahajan, L. L. Miller, 
	 T. Gu, A. I. Goldman, M. B. Maple, D. A. Gajewski, E. J. Freeman, N. R. Dilly, R. P. Dickey,
	 J. Merrin, K. Kojima, G. M. Luke, Y. J. Uemura, O. Chmaissem and J. D. Jorgensen:
	 Phys. Rev. Lett. \textbf{78} (1997) 3729.
 \bibitem{Urano2000} C. Urano, M. Nohara, S. Kondo, F. Sakai, H. Takagi, T. Shiraki and T. Okubo:
	 Phys. Rev. Lett. \textbf{85} (2000) 1052.
 \bibitem{Matsuno2001} K. Matsuno, T. Katsufuji, S. Mori, Y. Moritomo,
	 A. Machida, E. Nishibori, M. Takata, M. Sakata, N. Yamamoto and H. Takagi:
	 J. Phys. Soc. Jpn. \textbf{70} (2001) 1456.
 \bibitem{Matsuno2003} K. Matsuno, T. Katsufuji, S. Mori, M. Nohara, A. Machida, Y. Moritomo, 
	 K. Kato, E. Nishibori, M. Takata, M. Sakata, K. Kitazawa and H. Takagi: 
	 Phys. Rev. Lett. \textbf{90} (2003) 096404.
 \bibitem{Horibe2006} Y. Horibe, M. Shingu, K. Kurushima, H. Ishibashi,
	 N. Ikeda, K. Kato, Y. Motome, N. Furukawa, S. Mori and T. Katsufuji:
	 Phys. Rev. Lett. \textbf{96} (2006) 086406.
 \bibitem{Slater1954}  J. C. Slater and G. F. Koster: Phys. Rev. \textbf{94} (1954) 1498.
 \bibitem{Fujimori1988} A. Fujimori, K. Kawakami and N. Tsuda: Phys. Rev. B \textbf{38} (1988) 7889.
 \bibitem{Kanamori1963} J. Kanamori: Prog. Theor. Phys. \textbf{30} (1963) 275.
 \bibitem{Matsuno1999} J. Matsuno, A. Fujimori and L. F. Mattheiss: Phys. Rev. B \textbf{60} (1999) 1607.
 \bibitem{Altmann1994} Formally, the point group of the Hamiltonian is $ D_{3d} = D_3 \otimes C_i $.
 \bibitem{note} Another difficulty in the quantitative comparison is the spin quantum number
 in the magnetic excited state.
 The fitting of the temperature dependence of the magnetic susceptibility depends on
 the spin quantum number which is unknown experimentally at this stage.
 \bibitem{Lee2002} S.-H. Lee, C. Broholm, W. Ratcliff, G. Gasparovic, T. H. Kim, Q. Huang and S-W. Cheong:
	 Nature \textbf{418} (2002) 856.
 \bibitem{Khomskii2004} D. I. Khomskii and T. Mizokawa: Phys. Rev. Lett. \textbf{94} (2005) 156402.
 \bibitem{Pen1997} H. F. Pen, J. van den Brink, D. I. Khomskii and G. A. Sawatzky:
	 Phys. Rev. Lett. \textbf{78} (1997) 1323.
 \bibitem{Shimomura2004} Y. Shimomura, S. Miyahara and N. Furukawa: J. Phys. Soc. Jpn. \textbf{73} (2004) 1623.
\end{thebibliography}
\end{document}